\documentclass[letterpaper,12pt]{report}
\usepackage{fullpage}

\usepackage{amsmath, amsthm, amssymb}
\usepackage{graphicx}
\usepackage{verbatim}
\usepackage{fancyhdr}
\usepackage[table]{xcolor}
\usepackage[all]{xy}
\usepackage{tikz}
\usepackage{caption}
\usepackage{color}

\graphicspath{{Images/}}
 
\title{\textbf{On the Phase Space of Block-Hiding Strategies in Bitcoin-like networks}}

\author{Assaf Shomer\footnote{Contact the author at assafshomer at gmail dot com. Support this research with bitcoins (18wuRLYvbpei2EEUSykjms89cVciWBfcFF), or litecoins (LXT14HrbdQxU3cMiwuMPyEvp47frZkCD92).}\\
}

\date{\today}

\begin{document}

\maketitle

\begin{abstract}
We calculate the probability of success of block-hiding mining strategies in Bitcoin-like networks.
These strategies involve building a secret branch of the block-tree and publishing it opportunistically, aiming to replace the top of the main branch and rip the reward associated with the secretly mined blocks. We identify two types of block-hiding strategies and chart the parameter space where those are more beneficial than the standard mining strategy described in Nakamoto's paper.
Our analysis suggests a generalization of the notion of the relative hashing power as a measure for a miner's influence on the network. Block-hiding strategies are beneficial only when this measure of influence exceeds a certain threshold.

\end{abstract}
\pagenumbering{roman}
\tableofcontents
\chapter{Introduction}\label{chap:intro}
\pagenumbering{arabic}
Bitcoin is the world's first open source, decentralized digital currency. Bitcoin's main innovation is the ability to capture value in digital tokens through the creation of ``digital scarcity" \textit{independent of a central authority}. Clearly, a digital token that aims to capture value cannot achieve it's goal if it can be created or duplicated arbitrarily. Precious metals for instance, are scarce due to physical limitations. Fiat money is scarce through regulations and laws. All previous attempts to create a scarce digital resource depended upon a central authority that validates transactions and forces scarcity by disallowing arbitrary creation and multiple usage of the same token. 

Bitcoin achieves the same goals without a central authority. Transaction validation is achieved by duplicating the entire history of all transactions in all nodes of a distributed peer to peer computer network, thus allowing each node in the network to verify transaction validity independently. Each new transaction is immediately transmitted to the entire network. Because of the network's lack of centralization different nodes may be aware of a different, possibly conflicting set of transactions at a given time. 

Bitcoin's main innovation is a way to achieve consensus about \emph{accepted transaction history} amongst all nodes of the network. This is done by having each node independently bundle a set of valid transactions of his choice into a data structure called a \textit{block} in a process called ``mining". Mining a block demands computational effort, a.k.a \textit{proof of work}. Each valid block references it's predecessor thus creating a \textit{block-chain} which is the globally accepted history of transactions. Nodes that find valid blocks are rewarded with new bitcoins, creating an economic incentive to mining.

Once a valid block is found it is immediately transmitted to the network where all other nodes can easily (and quickly) verify it's validity and accept the one-block-longer block-chain as the new valid history of transactions. However, since block creation is random it is possible that more than one node manages to create a valid block and transmit it to the network. Each node chooses the longest branch\footnote{More precisely, the branch with maximal proof of work.} it is being made aware of first as the block-chain. When several branches are in state of a tie, the network's consensus about the ``true" block-chain is temporarily broken. Consensus is restored when a new block is found, breaking the tie, a fact that can be globally accepted. 

Thus, Bitcoin's consensus is subject to probability. A transaction that at one time is considered part of the globally accepted history of transactions may not be considered so in the future. However, as shown in \cite{Bitcoin}, the probability that a transaction included in the block-chain is later removed\footnote{Together with the block containing it} diminishes exponentially as more blocks are compounded on top of it. Another limitation to Bitcoin's consensus mechanism is that a node possessing enough computational power can ``hijack" the block-chain and dictate the global set of accepted transactions to be any desired set of valid transactions\footnote{Miners with more than $50\%$ of the networks resources can do that with $100\%$ probability of success. This is known as a the $51\%$ attack.}. These limitations to Bitcoin's consensus mechanism are the price being paid for not using a central authority. 

Bitcoin's consensus mechanism suffers from another weakness. The Bitcoin protocol assumes miners publish newly found blocks immediately and that every miner shifts it's mining effort to the top of the block-chain as soon as it is being made aware of a new block. However, these behaviours are not enforced and a necessary consequence of decentralization is that every miner is free to mine as she sees fit. If a miner ``breaks the rules" too aggressively, her blocks are in danger of being rejected by the network. However, some wiggle room exists for miners to participate without strictly following the Bitcoin protocol. For example, a miner can choose not to share a newly found block and build a secret branch of the block-tree that she only reveals opportunistically. Such non-traditional strategies were discussed in \cite{Selfish,Lear} and shown to enable miners to increase their profit. 

In this paper we are interested in finding out if and when non-standard mining strategies that involve building a secret branch of the block-tree and publishing it opportunistically (dubbed ``block-hiding" strategies) give miners a higher probability of success in mining blocks, compared to the standard strategy outlined in the Bitcoin protocol.

\chapter{Three mining strategies}\label{chap:strategies}

\section{The Standard Mining Strategy}
The standard mining strategy follow the bitcoin protocol described in \cite{Bitcoin}. Such miners publish each block as soon as it is discovered and switch their mining efforts to the head of the block-chain\footnote{In practice different miners may be aware of different branches of the block-tree at a given moment. Such differences are resolved with very high probability once a new block is found.} as soon as they become aware of a new valid block.

\begin{equation}\label{theblockchain}\nonumber
\dots\rightarrow\mathit{B}_L\rightarrow\mathit{B}_{L+1}\rightarrow\mathit{B}_{L+2}\rightarrow\dots
\end{equation}

\section{Block-Hiding Strategies}
Miners following this type of mining strategies do not share newly found blocks and instead work on extending a $\bf{secret}$ branch of the block-tree. The miners publish their secret branch when it is most beneficial to them. 

\begin{eqnarray}\label{fig:blockhidingchain}
 \dots \rightarrow \mathit{B}_L\rightarrow &\mathit{B}_{L+1}\rightarrow\mathit{B}_{L+2}
\rightarrow\dots\rightarrow\mathit{B}_{L+n} \qquad\qquad \mathrm{Main}\\\nonumber
\searrow & \\\nonumber
\qquad \qquad \qquad & \widetilde{\mathit{B}}_{L+1}\rightarrow\widetilde{\mathit{B}}_{L+2}
\longrightarrow \quad \dots \longrightarrow\widetilde{\mathit{B}}_{L+m}\qquad \mathrm{Secret}
\end{eqnarray}

\subsection{Type I (try to win)}
Type I miners mine a secret branch until it is \textit{longer than the main branch}. At this time they can publish it and replace the last $n$ blocks mined by the Standard miners in favour of their $m>n$ secretly mined ones.

\subsection{Type 0 (reach a tie and get some help)}
Type 0 miners mine a secret branch until it is of the  \textit{same length as the main branch}. At this time they publish it. Now the network is bifurcated. The Type 0 miners joined by some of the standard miners will mine on top of the newly published Type 0 branch. The rest of the standard miners continue working on the standard branch. If the former manage to find a new block first then the Type 0 strategy was successful. 

\section{Our Goal}\label{subsec:goal}

A miner of relative power $q$ can choose to follow any one of the mining strategies. If she selects to follow the gospel of \cite{Bitcoin} the probability of success in mining a new block equals $q$. Alternatively, she could follow one of the block-hiding strategies, and when the rest of the network mines a new block, mine a secret branch and publish it opportunistically instead of immediately switching her mining efforts to the head of the chain.  

Our goal in this work is to analyse \textit{which mining strategy is most beneficial in that it gives the highest probability of eventually claiming the rewards associated with the next block.} 
To that effect we calculate the probability that a block-hiding miner succeeds in replacing a block (and possibly some number of confirmation blocks on top of it) by publishing a secretly mined branch of the block-tree. Alternatively, this can be viewed as the probability that a block which is part of the block-chain will not survive (i.e. be part of the block-chain in the future) due to the effort of a block-hiding miner.

\chapter{Setup}\label{chap:calc}

Let us denote by $\mathcal{H}$ the total hashing power of the network and divide it abstractly into a \emph{Standard} part which holds a portion $p\mathcal{H}$ of the total hashing power (where $p \in [0,1]$) and a \emph{Block hiding} part, which holds the rest $q\mathcal{H}=(1-p)\mathcal{H}$. 

We start our analysis at a given point in time where the block-chain is of length $L$ and denote the last block mined as $\mathit{B}_L$. As time marches on the Standard miners continue to mine on top of it ($\mathit{B}_{L+1}, \mathit{B}_{L+2}, \dots$) while the block hiding miners are building a separate branch on top of $\mathit{B}_L$ ($\mathit{\tilde{B}}_{L+1}, \mathit{\tilde{B}}_{L+2}, \dots$). This is depicted in figure \ref{fig:blockhidingchain}. The block-hiding miners aim to replace the top of the chain mined on top of $\mathit{B}_L$ by using one of the two block-hiding strategies.

\section{Calculating the probability of success}

In order to calculate the probability that a block in the block-chain will be removed due to the effort of a block-hiding miner we first calculate the probability that the block-hiding miners manage to extend their secret branch on top of a certain block by $m$ blocks while the block-chain added $n$ confirmations on top of the same block (as depicted in figure \ref{fig:blockhidingchain}) and multiply by the probability that starting from such a configuration the block-hiding miner manages to catchup with the main chain. Our analysis follows the one presented in \cite{Doublespend}.

\subsection{Getting to the starting point}
Treating block mining as a negative binomial random variable, the probability $\mathit{P_{n,p}(m)}$ that $m$ blocks are mined in the secret branch {\bf before} $n$ blocks are mined in the main branch is proportional to $p^nq^m$ and can be shown
(appendix \ref{app:probmath}) to be given by

\begin{equation}\label{eq:pnm}
\mathit{P}_{n,q}(m)={n + m -1\choose m}(1-q)^nq^m \quad n=1,2,\dots
\end{equation}

\subsection{Catching up from the starting point}
The probability $\mathit{a}_{n,m}^{(r)}(q)$ that a block-hiding miner manages to catch-up and overtake the block-chain by at least $r$ blocks, given the situation above\footnote{Namely, that until the moment the main network mines it's $n$th block on top of $\mathit{B}_L$, the block-hiding miner manages to mine $m$ blocks on top of it.} is given by a Markov chain that depends only on the advantage $z=n-m$ of the main network over the block-hiding miner and the parameter $r$. Formally, the chain satisfies the recurrence relation
\begin{equation}\label{eq:markov}
\mathit{a}^{(r)}_z(q)=(1-q)\mathit{a}^{(r)}_{z+1}(q)+q\mathit{a}^{(r)}_{z-1}(q)
\end{equation}

with boundary conditions encoding the fact that a success is defined by the secret branch being longer than the main branch by at least $r$ blocks

\begin{equation}\label{eq:markovbc}
\texttt{Boundary Conditions:}
\begin{cases}
\mathit{a}^{(r)}_{-r}=1 & \\ 
\mathit{a}^{(r)}_{\infty}=0 & \\
\end{cases}
\end{equation}

The relation \ref{eq:markov} can be solved with boundary conditions \ref{eq:markovbc} by 

\begin{equation}\label{eq:az}
\mathit{a}^{(r)}_z(q)=\begin{cases}\left( \dfrac{q}{1-q}\right)^{z+r} & q\in [0,\frac{1}{2}] \quad \mathrm{and} \quad  -r<z\in\mathcal{Z} \\ \quad 1 & \mathrm{otherwise} \end{cases}
\end{equation}

\chapter{Type I Strategy}

Let $\mathit{Q}(q)$ be the probability that a Type I miner succeeds. By definition, the Type I strategy is successful when applied on top of $\mathit{B}_L$ if the miner manages to catch-up on $B_{L+1}$ and win by at least one block, after starting with $m=0,1,\dots$ secret blocks. 
By publishing the secret branch the miner replaces $\mathit{B}_{L+1}$ and any blocks mined by the network on top of it.

The starting point for the catch-up process for some $m$ is shown below:

\begin{eqnarray}\label{blockhidingboundary}
 \dots \rightarrow \mathit{B}_L\rightarrow &\mathit{B}_{L+1} \qquad\qquad\qquad\qquad\qquad\qquad\qquad\quad \mathrm{Main}\\\nonumber
\searrow & \\\nonumber
\qquad \qquad \qquad & \widetilde{\mathit{B}}_{L+1}\rightarrow\widetilde{\mathit{B}}_{L+2}
\longrightarrow \quad \dots \longrightarrow\widetilde{\mathit{B}}_{L+m}\qquad \mathrm{Secret}
\end{eqnarray}

By definition of the Type I strategy, the secret branch needs to be longer by at least one block\footnote{Hence the name: ``Type \textbf{I}".}, so we need to set the boundary condition in \ref{eq:markovbc} to $r=1$. 

\section{Block Revocation Probability}

The probability of success of a Type I miner is given by the a sum over all lengths of the secret branch at the beginning of catch-up. For each length $m$ we multiply the probability of getting to that starting point \ref{eq:pnm} while the main branch mines $\mathit{B}_{L+1}$ by the probability of catching up with the main branch from that starting point. The result is the following sum: 

\begin{equation}\label{eq:qofpdef}
\mathit{Q}(q)=\sum_{m=0}^{\infty}\mathit{P}_{1,q}(m)\mathit{a}^{(1)}_{1-m}(q)
\end{equation}

which gives (see details in appendix \ref{app:qofq})

\begin{equation}\label{eq:qofp}
\mathit{Q}(q)=
\begin{cases}
\dfrac{q^2}{1-q}\left(3-2q\right) & \quad q \in [0,\frac{1}{2}] \\
1 & \quad q \in [\frac{1}{2},1] 
\end{cases}
\end{equation}

In Figure \ref{fig:PlotProbOfSuccess} we plot the probability of a successful Type I strategy as a function of the relative hashing power $q$. As a reference we also plot the probability of success in mining a block for a standard miner with the same hashing power $q$.

\noindent%
\begin{minipage}{\linewidth}
\makebox[\linewidth]{%
  \includegraphics[keepaspectratio=true,scale=0.7]{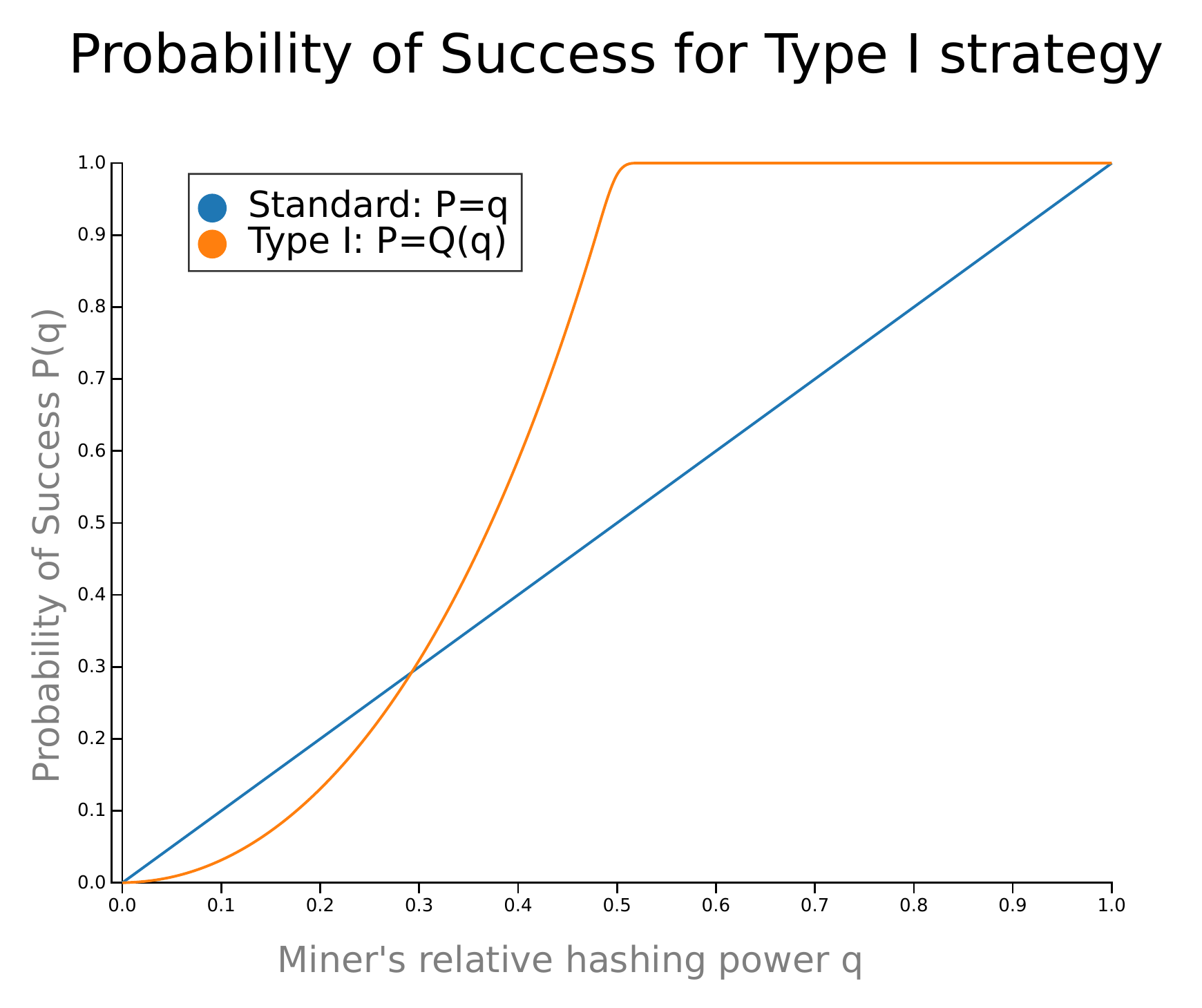}}
\captionof{figure}{
The orange curve plots the probability that a Type I miner manages to revoke the next block mined by the network and replace it (and any blocks mined on top of it) with blocks she mined in her secret branch. The blue curve is the baseline probability of harvesting a block by a Standard miner with the same hashing power $q$
}
\label{fig:PlotProbOfSuccess}
\end{minipage}

\section{Block Revocation after $n$ confirmations}

We can use equations \ref{eq:az} and \ref{eq:pnm} to calculate the probability of success of a Type I miner under the extra constraint that $\mathit{B}_{L+1}$ receives at least $n$ confirmation before being revoked (details of the calculation are given in appendix \ref{app:calcqofq})

\begin{eqnarray}\label{eq:qn}
&\mathit{Q}_{(n)}^{(1)}(q)=\sum_{m=0}^{\infty}\mathit{P}_{n,q}(m)\mathit{a}^{(1)}_{n-m}(q)=\\
&\begin{cases}
1-\sum_{m=0}^n{n + m -1\choose m}\left[(1-q)^nq^m-(1-q)^{m-1}q^{n+1}\right] & \quad q \in [0,\frac{1}{2}] \\
1 & \quad q \in [\frac{1}{2},1] 
\end{cases}
\end{eqnarray}

Below we show the probability of block-revocation by a Type I miner for some values of $q$ and some number of confirmations. If we use Satoshi's heuristics for safely considering a block as part of the block-chain if the probability of revocation by an attacker with $10\%$ of the hashing power is less than $0.1\%$ we see that $5$ confirmations will do. 

\begin{table}
\resizebox{18cm}{!} {
    \begin{tabular}{l|llllllllll}
    q  & n=1     & n=2     & n=3     & n=4     & n=5     & n=6     & n=7     & n=8     & n=9     & n=10    \\ \hline
    0\% & \colorbox{gray}{0.00\%}  & \colorbox{gray}{0.00\%}  & \colorbox{gray}{0.00\%}  & \colorbox{gray}{0.00\%}  & \colorbox{gray}{0.00\%}  & \colorbox{gray}{0.00\%}  & \colorbox{gray}{0.00\%}  & \colorbox{gray}{0.00\%}  & \colorbox{gray}{0.00\%}  & \colorbox{gray}{0.00\%}  \\
    2\% & 0.12\%  & \colorbox{gray}{0.01\%}  & \colorbox{gray}{0.00\%}  & \colorbox{gray}{0.00\%}  & \colorbox{gray}{0.00\%}  & \colorbox{gray}{0.00\%}  & \colorbox{gray}{0.00\%}  & \colorbox{gray}{0.00\%}  & \colorbox{gray}{0.00\%}  & \colorbox{gray}{0.00\%}  \\
    4\% & 0.49\%  & \colorbox{gray}{0.06\%}  & \colorbox{gray}{0.01\%}  & \colorbox{gray}{0.00\%}  & \colorbox{gray}{0.00\%}  & \colorbox{gray}{0.00\%}  & \colorbox{gray}{0.00\%}  & \colorbox{gray}{0.00\%}  & \colorbox{gray}{0.00\%}  & \colorbox{gray}{0.00\%}  \\
    6\% & 1.10\%  & 0.21\%  & \colorbox{gray}{0.04\%}  & \colorbox{gray}{0.01\%}  & \colorbox{gray}{0.00\%}  & \colorbox{gray}{0.00\%}  & \colorbox{gray}{0.00\%}  & \colorbox{gray}{0.00\%}  & \colorbox{gray}{0.00\%}  & \colorbox{gray}{0.00\%}  \\
    8\% & 1.98\%  & 0.49\%  & 0.13\%  & \colorbox{gray}{0.03\%}  & \colorbox{gray}{0.01\%}  & \colorbox{gray}{0.00\%}  & \colorbox{gray}{0.00\%}  & \colorbox{gray}{0.00\%}  & \colorbox{gray}{0.00\%}  & \colorbox{gray}{0.00\%}  \\
    10\% & 3.11\%  & 0.95\%  & 0.30\%  & 0.10\%  & \colorbox{gray}{0.03\%}  & \colorbox{gray}{0.01\%}  & \colorbox{gray}{0.00\%}  & \colorbox{gray}{0.00\%}  & \colorbox{gray}{0.00\%}  & \colorbox{gray}{0.00\%}  \\
    12\% & 4.52\%  & 1.63\%  & 0.61\%  & 0.23\%  & \colorbox{gray}{0.09\%}  & \colorbox{gray}{0.04\%}  & \colorbox{gray}{0.01\%}  & \colorbox{gray}{0.01\%}  & \colorbox{gray}{0.00\%}  & \colorbox{gray}{0.00\%}  \\
    14\% & 6.20\%  & 2.56\%  & 1.10\%  & 0.48\%  & 0.21\%  & 0.10\%  & \colorbox{gray}{0.04\%}  & \colorbox{gray}{0.02\%}  & \colorbox{gray}{0.01\%}  & \colorbox{gray}{0.00\%}  \\
    16\% & 8.17\%  & 3.78\%  & 1.82\%  & 0.89\%  & 0.44\%  & 0.22\%  & 0.11\%  & \colorbox{gray}{0.06\%}  & \colorbox{gray}{0.03\%}  & \colorbox{gray}{0.02\%}  \\
    18\% & 10.43\% & 5.33\%  & 2.82\%  & 1.52\%  & 0.84\%  & 0.46\%  & 0.26\%  & 0.15\%  & 0.08\%  & \colorbox{gray}{0.05\%}  \\
    20\% & 13.00\% & 7.24\%  & 4.17\%  & 2.45\%  & 1.46\%  & 0.88\%  & 0.53\%  & 0.32\%  & 0.20\%  & 0.12\%  \\
    22\% & 15.89\% & 9.54\%  & 5.92\%  & 3.74\%  & 2.39\%  & 1.54\%  & 1.00\%  & 0.66\%  & 0.43\%  & 0.28\%  \\
    24\% & 19.10\% & 12.27\% & 8.12\%  & 5.47\%  & 3.73\%  & 2.56\%  & 1.77\%  & 1.23\%  & 0.86\%  & 0.61\%  \\
    26\% & 22.66\% & 15.45\% & 10.83\% & 7.72\%  & 5.57\%  & 4.05\%  & 2.96\%  & 2.18\%  & 1.61\%  & 1.19\%  \\
    28\% & 26.57\% & 19.12\% & 14.11\% & 10.58\% & 8.01\%  & 6.12\%  & 4.70\%  & 3.63\%  & 2.81\%  & 2.18\%  \\
    30\% & 30.86\% & 23.30\% & 18.01\% & 14.12\% & 11.18\% & 8.91\%  & 7.15\%  & 5.75\%  & 4.65\%  & 3.77\%  \\
    32\% & 35.54\% & 28.02\% & 22.56\% & 18.41\% & 15.16\% & 12.56\% & 10.46\% & 8.75\%  & 7.34\%  & 6.18\%  \\
    34\% & 40.64\% & 33.31\% & 27.83\% & 23.53\% & 20.05\% & 17.19\% & 14.81\% & 12.80\% & 11.10\% & 9.65\%  \\
    36\% & 46.17\% & 39.20\% & 33.85\% & 29.54\% & 25.96\% & 22.93\% & 20.35\% & 18.11\% & 16.17\% & 14.46\% \\
    38\% & 52.17\% & 45.72\% & 40.66\% & 36.49\% & 32.95\% & 29.89\% & 27.21\% & 24.85\% & 22.74\% & 20.86\% \\
    40\% & 58.67\% & 52.91\% & 48.30\% & 44.43\% & 41.08\% & 38.14\% & 35.52\% & 33.16\% & 31.02\% & 29.06\% \\
    42\% & 65.69\% & 60.78\% & 56.80\% & 53.40\% & 50.41\% & 47.75\% & 45.34\% & 43.14\% & 41.11\% & 39.24\% \\
    44\% & 73.29\% & 69.39\% & 66.18\% & 63.42\% & 60.97\% & 58.75\% & 56.72\% & 54.85\% & 53.11\% & 51.47\% \\
    46\% & 81.51\% & 78.76\% & 76.49\% & 74.52\% & 72.75\% & 71.15\% & 69.66\% & 68.28\% & 66.98\% & 65.76\% \\
    48\% & 90.39\% & 88.95\% & 87.75\% & 86.71\% & 85.77\% & 84.91\% & 84.11\% & 83.37\% & 82.67\% & 82.00\% \\
    50\% & 100.00\% & 100.00\% & 100.00\% & 100.00\% & 100.00\% & 100.00\% & 100.00\% & 100.00\% & 100.00\% & 100.00\% \\
    ~   & ~       & ~       & ~       & ~       & ~       & ~       & ~       & ~       & ~       & ~       \\
    \end{tabular}
    }
    \caption {The probability of Block Revocation after n confirmations due to Type I mining. Gray cells denote probabilities smaller than $0.1\%$}
\end{table}

Adding the extra assumption that the Type I miner pre-mined a ``double-spend" block reproduces the results of \cite{Doublespend}:

\begin{eqnarray}\label{eq:ds}
&\sum_{m=0}^{\infty}\mathit{P}_{n,q}(m)\mathit{a}^{(1)}_{n-(m+1)}(q)=\\
&\begin{cases}
1-\sum_{m=0}^n{n + m -1\choose m}\left[(1-q)^nq^m-(1-q)^mq^n\right] & \quad q \in [0,\frac{1}{2}] \\
1 & \quad q \in [\frac{1}{2},1] 
\end{cases}
\end{eqnarray}

\chapter{Type 0 Strategy}

In this chapter we calculate the probability of success of a Type 0 mining strategy.
Instead of publishing the secret branch when it is longer than the main branch, a Type 0 miner publishes it one step before, when his secret branch is of the same length as the main branch (i.e. when they reach a tie). 

The reason this is potentially beneficial is that due to latency effects in the bitcoin network (recently discussed in \cite{Zoharetal}) not all miners share the same view of the entire block-tree at all moments. All Standard miners shift their efforts to the longest branch they know of, but for some period of time different parts of the network may be aware of different, and equally valid, longest branches. In such a case the network bifurcates. Each sub-network continues mining it's longest-branch until the next block is mined by either one and a new block-chain is established\footnote{In principle this type of block-chain bifurcation can continue to span multiple blocks, with exponentially decreasing probability.}. 
Following the notation used in \cite{Selfish} let us denote by $\gamma$ the ratio of standard miners that choose to mine on top of the newly-published-used-to-be-secret Type 0 branch.
This means that $q+\gamma p$ of the total hashing power is now dedicated to making the Type 0 branch the longest and with some probability this branch will end up as the winner. 

Let us denote the probability that this type of tie strategy succeeds by $S_{\gamma}(q)$.
We can calculate $S_{\gamma}(q)$ starting the same way as we did when we derived \ref{eq:qofp} but use a Markov chain with boundary condition reflecting a tie instead of wining. We then multiply that probability by the probability of catching up and wining the race with hashing power $q+\gamma p$, starting from that point.

Formally, we want to solve \ref{eq:markov} with boundary conditions $r=0$:

\begin{equation}\label{eq:bz}
\mathit{a}_z^{(0)}(q)=\begin{cases}\left( \dfrac{q}{1-q}\right)^z & q\in [0,\frac{1}{2}] \quad \mathrm{and} \quad z=0,1,2\dots \\ \quad 1 & \mathrm{otherwise} \end{cases}
\end{equation}

Using the same logic used to derive \ref{eq:qofpdef} we get the probability for a tie is given by

\begin{equation}\label{eq:qofpdef}
\mathit{T}(q)= \sum_{m=0}^{\infty}\mathit{P}_{1,q}(m)\mathit{a}^{(0)}_{1-m}(q)
\end{equation}

resulting in (see details in appendix \ref{app:tofq})

\begin{equation}\label{eq:qofp}
\mathit{T}(q)=
\begin{cases}
2q & \quad q \in [0,\frac{1}{2}] \\
1 & \quad q \in [\frac{1}{2},1] 
\end{cases}
\end{equation}

Now the Type 0 miner, \textbf{joined by $\gamma$ of the standard miners}, are competing with the rest of the standard miners. The probability to \textbf{win} starting from a tie is thus given by (see equation \ref{eq:az}) 

\begin{equation}\label{eq:azeroeff}
\mathit{a}^{(1)}_0(q_{eff})=\begin{cases} \dfrac{q_{eff}}{1-q_{eff}} & q_{eff}\in [0,\frac{1}{2}] \\  
\quad 1 & q_{eff}\in [\frac{1}{2},1] \\ \end{cases}
\end{equation}

where 
\begin{equation}\label{qeff}
q_{eff}=q+\gamma p=q+\gamma(1-q).
\end{equation}

The condition $q_{eff}\in [0,\frac{1}{2}]$ translates to 

\begin{equation}\label{qcrit}
0\leq q\leq q_{c}(\gamma)=\dfrac{1-2\gamma}{2-2\gamma}
\end{equation}

The curve $q_c(\gamma)$ (depicted in figure \ref{fig:qcritical}) satisfies $0 \leq   q_{c}(\gamma) \leq\dfrac{1}{2}$, monotonically decreases with $\gamma$ and hits $0$ when\footnote{$q_{eff}(\frac{1}{2})=\frac{1}{2}(1+q)$ which is bigger than $\frac{1}{2}$ for any $q$.} $\gamma=\frac{1}{2}$.

Based on all that, the solution to $S_\gamma(q)=T(q)\cdot a_0^{(1)}(q_{eff})$ breaks into three regimes:

\begin{equation}\label{eq:sofq}
S_\gamma(q)=\underbrace{T(q)}_{reach\; a\; tie}\cdot \underbrace{a_0^{(1)}(q_{eff})}_{win\;given\;a\;tie}=
\begin{cases}
2q\cdot\frac{q_{eff}}{1-q_{eff}}=2q\cdot\frac{q(1-\gamma)+\gamma}{(1-q)(1-\gamma)} & q\in [0,q_c] \\ 
2q & q\in [q_c,\frac{1}{2}] \\ 
1 & q\in [\frac{1}{2},1] \\ 
\end{cases}
\end{equation}

\noindent%
\begin{minipage}{\linewidth}
\makebox[\linewidth]{%
  \includegraphics[keepaspectratio=true,scale=0.4]{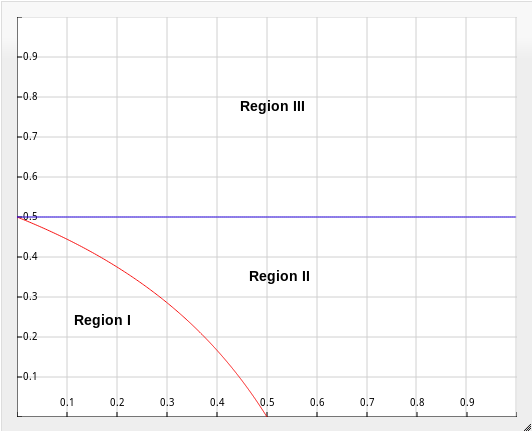}}
\captionof{figure}{
The three regions of the solution to $S_\gamma(q)$. The red curve is $q_c(\gamma)$. 
}
\label{fig:qcritical}
\end{minipage}
\linebreak

Note that if $\gamma\geq\frac{1}{2}$ the first regime does not exist and the solution degenerates to:

\begin{equation}\label{eq:sofq2}
S_{\gamma\geq\frac{1}{2}}(q)=T(q)\cdot a_0(q_{eff})=
\begin{cases}
2q & q\in [0,\frac{1}{2}] \\ 
1 & q\in [\frac{1}{2},1] \\ 
\end{cases} \quad = min(2q,1)
\end{equation}

In figure \ref{fig:type0pos} we plot the probability of success of the Type 0 strategy $S_\gamma(q)$ for various values of the parameter $\gamma$, side by side with the probabilities of success of the Standard and Type I strategies.

\noindent%
\begin{minipage}{\linewidth}
\makebox[\linewidth]{%
  \includegraphics[keepaspectratio=true,scale=0.8]{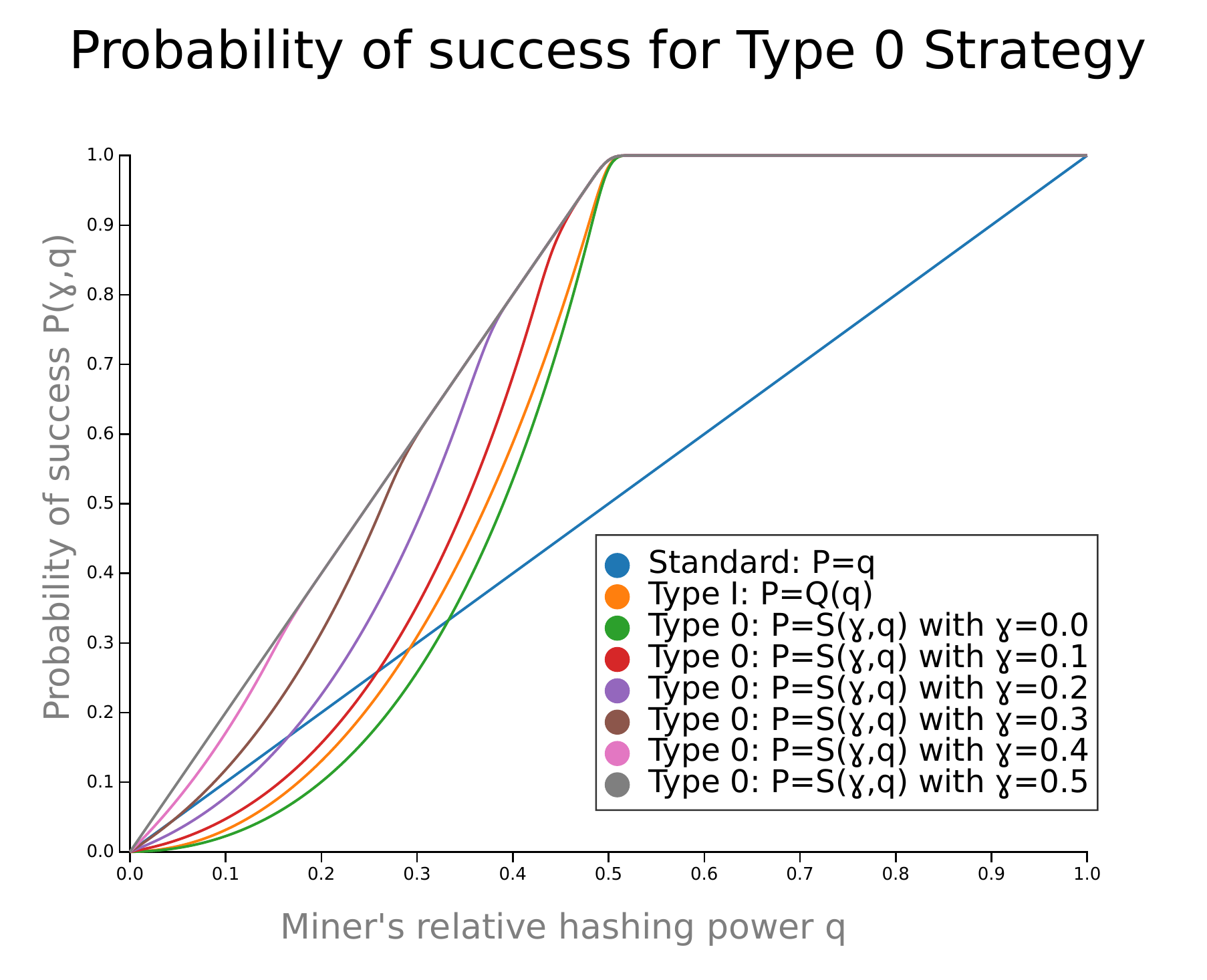}}
\captionof{figure}{The Type 0 strategy probability of success $S_\gamma(q)$, plotted for various values of the parameter $\gamma$ and against the probabilities for the Standard and Type I strategies.}
\label{fig:type0pos}
\end{minipage}
\linebreak

\chapter{The $\gamma,q$ phase space}\label{chap:gqphase}

Our aim in this chapter is to map the $\gamma,q$ phase space and find which strategy yields the maximal probability of success for the miner in each region.

\section{Type I vs. Standard}\label{TypeIoverStandard}
To find out how big $q$ needs to be for the Type I of strategy to become more beneficial than the standard strategy we need to solve for $Q(q)\geq q$

\begin{equation}\label{eq:type1overstandard}
\dfrac{q^2}{1-q}\left(3-2q\right) \geq q
\end{equation}

which since $0\leq q \leq 1$ gives the condition 

\begin{equation}\label{eq:qnot}
q\geq q_0=1-\frac{1}{\sqrt{2}} \sim 0.293
\end{equation}
We conclude that Type I strategy is better than the standard strategy for $q>q_0$. Once $q\geq \frac{1}{2}$ we get to the famous ``$51\%$ attack" where the Type I strategy is guaranteed to succeed, but even for $q_0\leq q \leq \frac{1}{2}$ Type I increases the probability of success for mining a new block compared to the standard strategy

\section{Type 0 vs. Standard}\label{Type0overStandard}

A Type 0 strategy is more beneficial than the standard strategy when $S_\gamma(q)\geq q$ where $S_\gamma(q)$ is given in equation \ref{eq:sofq}. 

In the second and third regimes of equation \ref{eq:sofq} (or for any $q$ if $\gamma\geq\frac{1}{2}$, see equation \ref{eq:sofq2} and figure \ref{fig:qcritical}) the Type 0 strategy is beneficial over the standard strategy for any $q$, because it is always true that $0\leq q \leq min(2q,1)$. 

In the first regime (i.e. when $q<q_c(\gamma)$) we can find at what value of $q$ the Type 0 strategy starts being more beneficial than the standard strategy by solving
\begin{equation}\label{eq:type0benefitonhonestequation}
2q\cdot\frac{q(1-\gamma)+\gamma}{(1-q)(1-\gamma)}\geq q
\end{equation}

which gives the condition $q_b(\gamma)\leq q \leq q_c(\gamma)$, where 

\begin{equation}\label{eq:qb}
q_b= \dfrac{1-3\gamma}{3-3\gamma}
\end{equation}

The curve $q_b(\gamma)$ designating the boundary where the Type 0 strategy starts becoming more beneficial than the standard strategy is plotted in Figure \ref{fig:qbqc}.
Note that if $\gamma=0$ this strategy is beneficial only when $q>\frac{1}{3}$ and if $\gamma\geq\frac{1}{3}$ it is beneficial for all $q$.

\noindent%
\begin{minipage}{\linewidth}
\makebox[\linewidth]{%
  \includegraphics[keepaspectratio=true,scale=0.6]{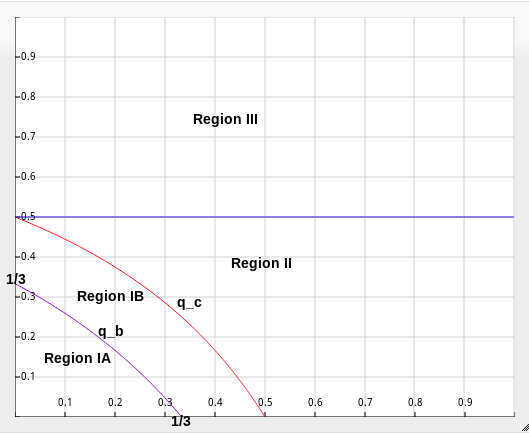}}
\captionof{figure}{The Type 0 strategy is more beneficial than the standard strategy outside of the inner curve $q_b(\gamma)$. Standard mining is more beneficial only in region IA.}
\label{fig:qbqc}
\end{minipage}
\linebreak

Taking all three regimes into account we conclude that the Type 0 strategy is more beneficial than the Standard strategy when

\begin{equation}\label{eq:0overhonest}
S_{\gamma}(q)\geq q \Longrightarrow
\begin{cases}
q>\dfrac{1-3\gamma}{3-3\gamma} & \gamma\in [0,\frac{1}{3}] \\ 
\mathit{any\;} q & \gamma\in [\frac{1}{3},1] \\ 
\end{cases}
\end{equation}

or, summarized further by 

\begin{equation}\label{eq:0overhonestsummary}
S_{\gamma}(q)\geq q \Longrightarrow
q>max\left\lbrace\dfrac{1-3\gamma}{3-3\gamma},0\right\rbrace
\end{equation}

To decide if Type 0 is beneficial or not it is not enough to compare it to the Standard strategy. Even in the regime where it is more beneficial than the Standard strategy we must compare it to Type I to decide which block-hiding strategy wins. In the next section we do exactly that and compare the two block-hiding strategies.

\section{Comparing Type 0 to Type I}

There are two interesting comparisons one can make between Type 0 and Type I strategies.
One is to compare how they match against the standard strategy,namely, for a given $\gamma$ do we first hit the regime where a Type 0 or a Type I strategy is more beneficial than the standard strategy.
The other is to ignore the standard strategy and ask, for a given value of $\gamma, q$ which is better, Type 0 or Type I.

\subsection{Which block-hiding is beneficial first}
Let us first tackle the first question and find out for a given $\gamma$ whether Type 0 or Type I wins first.
For $\gamma\geq\frac{1}{3}$, Type 0 wins already at $q=0$, so the interesting part is where $\gamma < \frac{1}{3}$ where we can compare $q_b(\gamma)$ (given in equation \ref{eq:qb}) with $q_0$ (given in equation \ref{eq:qnot}). Solving for the intersectin of the two curves

\begin{equation}\label{eq:qbornot}
\dfrac{1-3\gamma}{3-3\gamma}=1-\dfrac{1}{\sqrt{2}}
\end{equation}

We get a single intersection at a special value of $\gamma$

\begin{equation}\label{gamma0before1}
\gamma_c=1-\frac{2}{3}\sqrt{2}\sim 0.0572
\end{equation}

A Type 0 strategy is more beneficial than the standard strategy sooner (i.e. smaller $q$) than Type I for $\gamma \geq\gamma_c$.

Indeed, you can see in Figure \ref{fig:type0pos} that the green curve representing $\gamma=0$ lies below the orange curve which represents the Type I strategy, while the red curve representing $\gamma=0.1>\gamma_c$ lies above it.

To summarize, when $\gamma\geq\frac{1}{3}$ the Type 0 strategy is beneficial over the standard strategy for any value of $q$. When $\gamma<\frac{1}{3}$ , the hashing power of the block-hiding miner needs to exceed a threshold before a block-hiding strategy is beneficial. If $\gamma_c\leq\gamma\leq\frac{1}{3}$ we bump into the Type 0 first (the threshold given by $q_b=\frac{1-3\gamma}{3-3\gamma}$), while if $\gamma<\gamma_c$ we bump into Type I first (the threshold is given by $q_0=1-\frac{1}{\sqrt{2}}$).

\subsection{Type 0 vs. Type I}

Finally, ignoring the standard strategy for a moment, we can ask for the range of parameters $q,\gamma$ where the Type 0 strategy is more beneficial than the Type I strategy. Formally we need to solve:

\begin{equation}\label{eq:type0overtype1}
2q\cdot\frac{q(1-\gamma)+\gamma}{(1-q)(1-\gamma)}\geq \dfrac{q^2}{1-q}\left(
3-2q
\right)
\end{equation}

which gives the condition

\begin{equation}\label{eq:type0over1condition}
2q^2-q+\frac{2\gamma}{1-\gamma}\geq 0
\end{equation}

This condition is satisfied in two regimes for $\gamma$.
\begin{equation}\label{eq:type0over1gammaregimes}
S_{\gamma}(q)\geq Q(q) \Longrightarrow
\begin{cases}
\mathit{any\;} q & \gamma\in [\frac{1}{17},1] \\ 
q<q_-(\gamma)\quad \mathit{or}\quad q>q_+(\gamma) & \gamma\in [0,\frac{1}{17}] \\ 
\end{cases}
\end{equation}

where 

\begin{equation}\label{qplusminus}
q_{\pm}(\gamma)=\frac{1}{4}\left(1\pm\sqrt{\frac{1-17\gamma}{1-\gamma}}\right)
\end{equation}

\section{The Strategy Phase Space}\label{sec:phasespace}

We can chart the strategy phase space parametrized by $\gamma,q\in [0,1]^2$, and divide it into regions characterized by the most beneficial mining strategy: \textbf{Standard}, \textbf{Type 0} or \textbf{Type I}.

The $\gamma, q$ phase space is governed by four functions (really three intersecting curves):
\begin{itemize}
\item $q_0= 1-\frac{1}{\sqrt{2}}$ determining for what $q$ Type I is better than standard.
\item $max\lbrace q_b(\gamma)=\frac{1-3\gamma}{3-3\gamma},0\rbrace$ determining for what $q$ Type 0 is better than standard.
\item $q_+(\gamma)=\frac{1}{4}\left(1+\sqrt{\frac{1-17\gamma}{1-\gamma}}\right)$ 
\item $q_-(\gamma)=\frac{1}{4}\left(1-\sqrt{\frac{1-17\gamma}{1-\gamma}}\right)$
\end{itemize}
where the last two determine which strategy is better, Type 0 or I when $\gamma<\frac{1}{17}$.
Interestingly enough, the 3 functions $q_0, q_b(\gamma), q_+(\gamma)$ intersect in a single point $\gamma=\gamma_c, q=q_0$ which simplifies the structure of the phase space diagram, slicing it into exactly 6 regions each characterized by one of the 6 possible orderings between the 3 available strategies.

\begin{itemize}
\item The circular curve (created by the two branches $q_{\pm}$) determines, for a given $\gamma$, which of the two block-hiding strategies, Type 0 or Type I is more beneficial. Inside the circular region (and all the way to the $q$ axis) is the region where Type I is better than type 0. Outside this region Type 0 is better than Type 1. This is determined by equation \ref{eq:type0over1condition}. Note that this division doesn't specify whether any of the strategies is better than the standard one.
\item Type I strategy is more beneficial than the Standard strategy in the region above the horizontal line $q=q_0$. .
\item Type 0 strategy is more beneficial than the Standard strategy in the region above the monotonically decreasing curve $q_b(\gamma)$ (extending from $\frac{1}{3}$ on the $q$ axis to $\frac{1}{3}$ on the $\gamma$ axis and the continuing on the $\gamma$ axis all the way to $\gamma=1$).
\end{itemize}

\noindent%
\begin{minipage}{\linewidth}
\makebox[\linewidth]{%
  \includegraphics[keepaspectratio=true,scale=0.7]{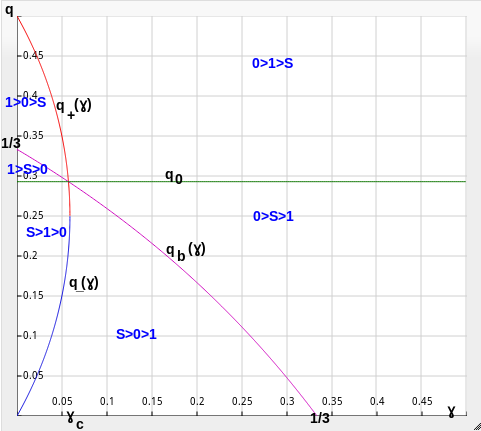}}
\captionof{figure}{The 3 curves sectioning the $\gamma, q$ phase space into 6 regions characterized by a hierarchy of the 3 strategies. For brevity we denote the Type I strategy by ``1", the Type 0 strategy by ``0" and the Standard strategy by ``S". For example, in the top region Type 0 is more beneficial than Type I which in turn is more beneficial than the standard strategy.}
\label{fig:fourfunctions}
\end{minipage}
\linebreak

A miner, seeking to maximize profit, can select the most beneficial strategy in each region. The resulting phase space is divided into 3 regions characterized by the winning strategy, is depicted in figure \ref{fig:qgammaphasespace}.

\noindent%
\begin{minipage}{\linewidth}
\makebox[\linewidth]{%
  \includegraphics[keepaspectratio=true,scale=0.7]{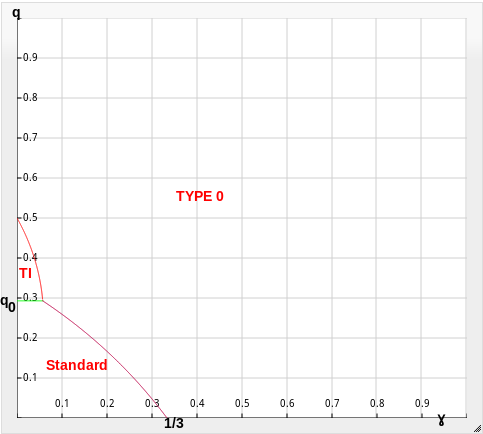}}
\captionof{figure}{The 3 regions of the $q,\gamma$ phase space. The standard strategy is best in the region near the origin. Type I is best in the little area on top of the standard region. In the rest of phase space the Type 0 strategy is most beneficial.}
\label{fig:qgammaphasespace}
\end{minipage}
\linebreak

\section{Comments}

Note that the regime where either block-hiding strategies are more beneficial than the standard strategy is bounded away from the origin. It is tempting to look at the radial distance from the origin of phase space as a measure of a ``miner's \textit{influence}" 

\begin{equation}
\mathtt{I}(q,\gamma)\equiv \sqrt{q^2+\gamma^2}.
\end{equation}

This definition is motivated by the rough symmetry\footnote{All we mean by that is that figure \ref{fig:qgammaphasespace} is almost symmetric under a rotation along the $45^{\circ}$ angle that rotates $\gamma \leftrightarrow q$.} between the parameters $\gamma$ and $q$.

There are a few remarks in order:
\begin{itemize}
\item Figure \ref{fig:qgammaphasespace} marks the regions where a block-hiding strategy in beneficial, but does not guarantee success. Success of either strategies is still guaranteed (i.e. the probability of success is strictly $1$) only in the top half of phase space, in the region where $q\geq \frac{1}{2}$ (the infamous $51\%$ attack).
\item The authors of \cite{Selfish} identified a region delimited by the curve $\frac{1-\gamma}{3-2\gamma}\leq q \leq \frac{1}{3}$ where the ``selfish" mining strategy is more beneficial than the standard one. As one would expect based on the fact that the ``selfish'' strategy utilizes a combination of the two strategies discussed here, this curve intersects both Type I and Type 0 regions as depicted in figure \ref{fig:selfish}.
\end{itemize}

\noindent%
\begin{minipage}{\linewidth}
\makebox[\linewidth]{%
  \includegraphics[keepaspectratio=true,scale=0.7]{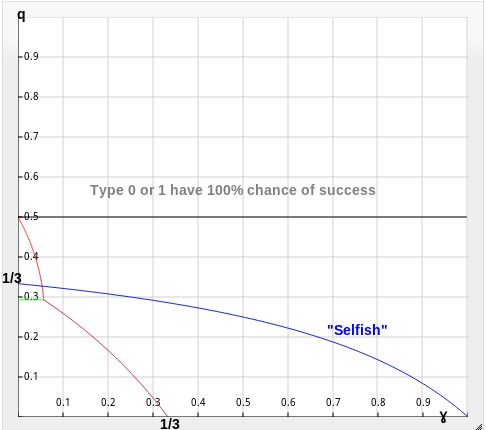}}
\captionof{figure}{In this plot we see the location of the curve describing the ``seflish" strategy of \cite{Selfish} within the $\gamma, q$ phase space.}
\label{fig:selfish}
\end{minipage}
\linebreak

\chapter{Summary and Conclusions}
The Bitcoin protocol leaves some aspects of mining as heuristics that are not enforced. This wiggle room allows miners to explore mining strategies that deviate from what was described in \cite{Bitcoin} but are still technically valid and will be accepted by all nodes in the Bitcoin network. Bitcoin's  main innovation is the ability to withstand ``Double Spend" attacks, which can be viewed as a particular case of a block-hiding mining strategy. Other variants were known for quite a while \cite{Lear,Selfish} and serve as an interesting theoretical ground for studying the limits of the Bitcoin protocol. 
 
In this paper we analysed the probability of success of block-hiding miners, charted the phase space  associated with block-hiding strategies and identified the most beneficial mining strategy in each regime. As a corollary we calculated the probability of block revocation due to the presence of a Type I miner. Our analysis extend the results of \cite{Selfish} and refine the notion of when exactly a miner becomes ``too big", starting to pose a potential threat to the Bitcoin network (see \cite{Centralization} for a recent discussion). 

As long as miners posses a small enough fraction of the total hashing power and are not too well connected to lure an disproportionally big part of the network to mine on top of their block in case of a tie, the Standard mining strategy as presented in \cite{Bitcoin} turns out to be the most beneficial. It would be interesting if keeping miners in the safe zone can be actively discouraged by additions to the protocol.

\section{Acknowledgements}
We would like to thank Aviv Zohar and Meni Rosenfeld for comments on earlier versions of this manuscript.

\appendix
\chapter{Calculation Details}
\section{Probability distribution} \label{app:probmath}
\begin{equation}\nonumber
\sum_{m=0}^{\infty}\mathit{P}_{n,q}(m)=p^n\sum_{m=0}^{\infty}{n + m -1\choose m}q^m=p^n\dfrac{1}{(1-q)^n}=1
\end{equation}

where we used the binomial identity holding for any complex $s$ inside the unit circle ($|s|<1$)
\begin{equation}\label{eq:binident}\nonumber
\dfrac{1}{(1-s)^n}=\sum_{k=0}^{\infty} {n + k -1\choose k}s^k
\end{equation}

\section{$\mathit{Q_{(n)}^{(r)}}(q)$ } \label{app:calcqofq} 

\begin{eqnarray}\label{eq:qngeneral}\nonumber
&\mathit{Q}_{(n)}^{(r)}(q)=\sum_{m=0}^{\infty}\mathit{P}_{n,q}(m)\mathit{a}^{(r)}_{n-m}(q)=\\ \nonumber
&
\begin{cases}
\sum_{m=0}^{n+r-1}{n + m -1\choose m}(1-q)^nq^m\left(\dfrac{q}{1-q}\right)^{n-m+r}+ \sum_{m=n+r}^{\infty}{n + m -1\choose m}(1-q)^nq^m & \quad q \in [0,\frac{1}{2}] \\ 
1 & \quad q \in [\frac{1}{2},1] 
\end{cases} \\ \nonumber
&=
\begin{cases}
\sum_{m=0}^{n+r-1}{n + m -1\choose m}(1-q)^nq^m\left[\left(\dfrac{q}{1-q}\right)^{n-m+r}-1\right]+ 1 & \quad q \in [0,\frac{1}{2}] \\
1 & \quad q \in [\frac{1}{2},1] 
\end{cases} \\ \nonumber
&=
\begin{cases}
\sum_{m=0}^{n+r-1}{n + m -1\choose m}\left[(1-q)^{m-r}q^{n+r} - (1-q)^nq^m\right]+ 1 & \quad q \in [0,\frac{1}{2}] \\
1 & \quad q \in [\frac{1}{2},1] 
\end{cases} \\ \nonumber
&=
\begin{cases}
1-\sum_{m=0}^{n+r-1}{n + m -1\choose m}\left[(1-q)^nq^m-(1-q)^{m-r}q^{n+r}\right] & \quad q \in [0,\frac{1}{2}] \\
1 & \quad q \in [\frac{1}{2},1] 
\end{cases} \\ \nonumber
\end{eqnarray}

\subsection{$\mathit{Q}(q)$}\label{app:qofq}
In the case $\mathit{Q}(q)\equiv\mathit{Q_{(1)}^{(1)}}(q)$ we get:

\begin{eqnarray}\label{eq:qnspecific}\nonumber
&\mathit{Q}(q)=\sum_{m=0}^{\infty}\mathit{P}_{1,q}(m)\mathit{a}^{(1)}_{1-m}(q)\\ \nonumber
&=
\begin{cases}
1-\sum_{m=0}^1\left[(1-q)q^m-(1-q)^{m-1}q^2\right] & \quad q \in [0,\frac{1}{2}] \\
1 & \quad q \in [\frac{1}{2},1] 
\end{cases} \\ \nonumber
&=
\begin{cases}
1-(1-q)+\dfrac{q^2}{1-q}-(1-q)q+q^2 & \quad q \in [0,\frac{1}{2}] \\
1 & \quad q \in [\frac{1}{2},1] 
\end{cases} \\ \nonumber
&=
\begin{cases}
\dfrac{q^2}{1-q}+2q^2 & \quad q \in [0,\frac{1}{2}] \\
1 & \quad q \in [\frac{1}{2},1] 
\end{cases} =
\begin{cases}
\dfrac{q^2}{1-q}\left(3-2q\right) & \quad q \in [0,\frac{1}{2}] \\
1 & \quad q \in [\frac{1}{2},1] 
\end{cases} \\ \nonumber
\end{eqnarray}

\subsection{$\mathit{T}(q)$}\label{app:tofq}
In the case $\mathit{T}(q)\equiv\mathit{Q_{(1)}^{(0)}}(q)$ we get:

\begin{eqnarray}\label{eq:qnspecific}\nonumber
&\mathit{T}(q)=\sum_{m=0}^{\infty}\mathit{P}_{1,q}(m)\mathit{a}^{(0)}_{1-m}(q)\\ \nonumber
&=
\begin{cases}
1-\left[(1-q)-q\right] & \quad q \in [0,\frac{1}{2}] \\
1 & \quad q \in [\frac{1}{2},1] 
\end{cases} =
\begin{cases}
2q & \quad q \in [0,\frac{1}{2}] \\
1 & \quad q \in [\frac{1}{2},1] 
\end{cases} \\ \nonumber
\end{eqnarray}

\newpage
\bibliographystyle{plain} % IEEEtr, plain, abbrv, alpha
\addcontentsline{toc}{chapter}{Bibliography}
\bibliography{blocksurvival}
\end{document}